\documentclass[a4paper,reqno]{amsart}
\usepackage[margin=2.5cm,papersize={19.75cm,28cm}]{geometry}
\usepackage[all]{xy}           
\usepackage{amssymb}           
\usepackage{hyperref}
\usepackage{eucal}
\usepackage{epsfig}
\usepackage{graphicx}

\numberwithin{equation}{section}

\newtheorem{definition}{Definition}[section]
\newtheorem{lemma}[definition]{Lemma}
\newtheorem{theorem}[definition]{Theorem}
\newtheorem{proposition}[definition]{Proposition}

\newtheorem{remarkth}[definition]{Remark}
\newtheorem{example}[definition]{Example}
\newenvironment{remark}{\begin{remarkth}\upshape}{\hfill$\diamond$\end{remarkth}}

\renewcommand{\emph}[1]{{\bfseries\itshape{#1}}}

\newcommand{\R}{\mathbb{R}}      
\newcommand{\N}{\mathbb{N}}      

\def\ra{\rightarrow}

\makeatother

\newcommand{\real}{\mathbb{R}}

\newcommand{\Su}{\mathbb{S}}

\begin{document}

\title[Optimal Control of Underactuated Mechanical Systems]{Optimal Control of Underactuated Mechanical Systems:\\ A Geometric Approach}

\author[L. Colombo]{Leonardo Colombo}
\address{L. Colombo:
Instituto de Ciencias Matem\'aticas  (CSIC-UAM-UC3M-UCM), Serrano 123, 28006 Madrid,
Spain and Departamento de Matem\'atica, Universidad Nacional de La Plata, Calle 50 y 115, La Plata, Buenos Aires,
Argentina} \email{leonardocolombo867@gmail.com}

\author[D.\ Mart\'{\i}n de Diego]{David Mart\'{\i}n de Diego}
\address{D.\ Mart\'{\i}n de Diego: Instituto de Ciencias Matem\'aticas (CSIC-UAM-UC3M-UCM), Serrano 123, 28006
Madrid, Spain} \email{d.martin@imaff.cfmac.csic.es}

\author[M. Zuccalli]{Marcela Zuccalli}
\address{M. Zuccalli:
Departamento de Matem\'atica, Universidad Nacional de La Plata, Calle 50 y 115, La Plata, Buenos Aires,
Argentina} \email{marce@mate.unlp.edu.ar}


\begin{abstract} In this paper, we consider a geometric formalism  for optimal control of underactuated mechanical systems. Our techniques are an adaptation of the classical Skinner and Rusk approach for the case of   Lagrangian dynamics with higher-order constraints.
We study  a regular case where it is possible to establish a symplectic framework and, as a consequence, to obtain a unique vector field determining the dynamics of the optimal control problem. These developments will allow us to develop a new class of geometric integrators based on   discrete variational calculus.

\end{abstract}

\thanks{This work has been partially supported by MEC (Spain)
 MTM 2007-62478, project ``Ingenio
Mathematica" (i-MATH) No. CSD 2006-00032 (Consolider-Ingenio 2010)
and S-0505/ESP/0158 of the Comunidad de Madrid.  L. Colombo  also wants to thank
CSIC for a JAE-intro grant. The authors wish to thank Sebasti\'an Ferraro,  \'Angel Castro and Germ\'an Zorba for
helpful comments }

\maketitle

\section{Introduction}
The mathematical activity in the last century in dynamical systems, mechanics and related areas has been extraordinary. The number of
applications has grown exponentially  and both basic science as well as several engineering
technologies are
profiting from this development. In the 1960s more sophisticated and powerful techniques coming from modern differential
geometry and topology have been introduced in their study, experiencing  a spectacular growth
in the last 50 years. Control and optimal control of mechanical systems has not ignored these developments, becoming now  a principal research focus of nonlinear control theory. In particular, there are an increasing interest in the control of underactuated mechanical systems (see \cite{bullolewis,cortes}).
These type of mechanical systems are  characterized by the fact that there are more degrees of freedom than actuators. This type of system is quite different from a mathematical and engineering perspective than fully actuated control systems where all the  degrees of freedom are actuated.

The class of underactuated mechanical systems are
abundant in real life for different reasons, for instance, as a result of design choices motivated by the search of less cost engineering devices or as a result of a failure  regime in fully actuated mechanical systems. The underactuated systems include spacecraft,
underwater vehicles, mobile robots, helicopters, wheeled vehicles, mobile robots, underactuated
manipulators...

On the other hand, there are  many papers in which optimal control problems are addressed using geometric techniques (see, for instance,  \cite{block2,Vu,Vu2,Su} and references therein). Now, we introduce an optimization strategy in an underactuated mechanical system, that is, we are interested in studying the implementation of devices in which a controlled
quantity is used to influence the behavior of the undeactuated system in order to achieve a desired goal (control)
using the most economical strategy (optimization).
Thus, in our  paper we develop a new geometric setting for optimal control of underatuated lagrangian systems  strongly inspired on the Skinner and Rusk formulation for singular Lagrangians systems~\cite{Skinner-Rusk}.
Since in this setting the controlled Euler-Lagrange equation are second-order differential equations we will need to implement an higher-order version of this classical Skinner and Rusk formalism \cite{hidden}.  This geometric procedure gives us an intrinsic version of the differential equations for optimal trajectories and permits us to detect the preservation of geometric properties (symplecticity, preservation of the hamiltonian, etc.).
For expository simplicity, we restrict ourselves in Section \ref{section5} to the so-called optimal control of superarticulated mechanical systems, in which only some of the degrees of freedom are controlled directly, with the remaining variables freely evolving subject only to dynamic interactions with the actuated degrees of freedom (see \cite{Ba,SeBa}). Obviously, our theory can be easily extended to more general class of underactuated lagrangian systems.

Moreover,  using a  discrete version of this variational
approach to optimal control of underactuated lagrangian systems  it is possible  to construct
discretization schemes for this type of systems
which preserve a discrete symplectic form. All
this type of geometric integrators has demonstrated, in worked examples,
an exceptionally good longtime behavior and obviously this
research is of great interest from numerical and geometric
considerations (see \cite{Hair}).

The paper is organized as follows. In Section \ref{section2} we
recall some geometric constructions and properties of higher-order
tangent bundles to make the paper self-contained. In section
\ref{section3}, we derive the main geometric objects of higher-order
variational calculus with higher-order constraints. In Section
\ref{section4}, we deduce the geometric framework for higher-order
mechanics using the Skinner and Rusk formalism. In Section \ref{section5} we apply the previous techniques to the case of optimal control of underactuated Lagrangian systems. A concrete example: the Cart-Pole system is carefully analyzed.  Finally, in Section \ref{section6} we derive the corresponding discrete version obtaining  a
numerical integrator which inherits some of the previously analyzed geometric
properties of the continuous optimal control problem for the underactuated system (symplecticity, momentum
preservation...).


\section{Higher-order tangent bundles} \label{section2}

In this section we recall some basic facts of the higher-order
tangent bundles theory. For more details see \cite{CSC,LR1}.

Let $Q$ be a  manifold of dimension $n$. An equivalence relation
is introduced in the set $C^{\infty}(\R, Q)$ of differentiable
curves from $\R$ to $Q$. By definition,  two given curves in $Q$
$\gamma_1(t)$ and $\gamma_2(t)$
where $t\in (-a, a)$ with $a\in \R$
have contact of order  $k$ at $q_0 = \gamma_1(0) = \gamma_2(0)$ if
there is a local chart $(\varphi, U)$ of $Q$ such that $q_0 \in U$
and
$$\frac{d^s}{dt^s}\left(\varphi \circ \gamma_1(t)\right){\Big{|}}_{t=0} =
\frac{d^s}{dt^s} \left(\varphi
\circ\gamma_2(t)\right){\Big{|}}_{t=0}\; ,$$ for $s = 0,...,k.$ This
is a well defined equivalence relation
in $C^{\infty}(\R,Q)$
and the equivalence class of a  curve $\gamma$ will be denoted by
$[\gamma ]_0^{(k)}.$ The set of equivalence classes will be
denoted by $T^{(k)}Q$
and we can see that
it is a differentiable manifold. Moreover, $ \tau_Q^k  : T^{(k)} Q
\rightarrow Q$ where $\tau_Q^k \left([\gamma]_0^{(k)}\right) =
\gamma(0)$ is a fiber bundle called the {\em tangent bundle of
order $k$} of $Q.$

We also  may
define the surjective mappings $\tau_Q^{(l,k)} : T^{(k)} Q
\rightarrow T^{(l)} Q,$ for $l\leq k$, given by
$\tau_Q^{(l,k)}\left([\gamma]_0^{(k)}\right) = [\gamma]_0^{(l)}.$
It is easy to see that $T^{(1)} Q \equiv TQ$, the tangent bundle
of $Q$, $T^{(0)} Q \equiv Q$ and $\tau_Q^{(0,k)}=\tau_Q^k$.

Given a differentiable function $f: Q\longrightarrow \R$ and $l
\in \{0,...,k\}$, its $l$-lift $f^{(l, k)}$ to $T^{(k)}Q$, $0\leq
l\leq k$, is the
differentiable
function defined as
\[
f^{(l, k)}([\gamma]^{(k)}_0)=\frac{d^l}{dt^l}
\left(f \circ \gamma(t)\right){\Big{|}}_{t=0}\; .
\]
Of course, these definitions can be applied to
functions defined
on open sets of $Q$.

From a local chart $(q^i)$ on a neighborhood $U$ of $Q$, it is possible to induce local coordinates
 $(q^{(0)i},q^{(1)i},\dots,q^{(k)i})$ on
$T^{(k)}U=(\tau_Q^k)^{-1}(U)$, where $q^{(s)i}=(q^i)^{(s,k)}$ if
$0\leq s\leq k$. Sometimes, we will use the standard  conventions, $q^{(0)i}\equiv q^i$, $q^{(1)i}\equiv \dot{q}^i$ and
$q^{(2)i}\equiv \ddot{q}^i$.

Given a vector field $X$ on $Q$, we define its $k$-lift $X^{(k)}$
to $T^{(k)}Q$ as the unique vector field on $T^{(k)}Q$ satisfying the following
identities
\[
X^{(k)}(f^{(l, k)})=(X(f))^{(l,
k)}\, , \hspace{1.2cm} 0\leq l\leq k\, ,
\]
for all differentiable function $f$ on $Q$.
 In coordinates, the $k$-lift of a
vector field $\displaystyle{X=X^i\frac{\partial}{\partial q^i}}$
is
\[
X^{(k)}=(X^i)^{(s, k)}\frac{\partial}{\partial
q^{(s)i}}\; .
\]

Now, we consider the canonical immersion
$j_k: T^{(k)}Q\rightarrow T(T^{(k-1)} Q)$ defined as
$j_k([\gamma]_0^{(k)})=[{\gamma}^{(k-1)}]_0^{(1)}$, where
${\gamma}^{(k-1)}$ is the lift of the curve
$\gamma$ to $T^{(k-1)}Q$; that is,  the curve ${\gamma}^{(k-1)}: \R\rightarrow
T^{(k-1)}Q$ is given by $\gamma^{(k-1)}(t)=[\gamma_t]_0^{(k-1)}$
where $\gamma_t(s)=\gamma(t+s)$. In local coordinates
$$
j_k(q^{(0)i},q^{(1)i},q^{(2)i},...q^{(k)i})=(q^{(0)i},q^{(1)i},\dots,q^{(k-1)i};q^{(1)i},
q^{(2)i},\dots,q^{(k)i})\; .
$$

We use the map $j_{k}$ to construct the differential operator
$d_{T}$ which maps a function $f$ on $T^{(k)}Q$ into a function
$d_{T}f$ on $T^{(k+1)}Q$

$$d_{T}f([\gamma]_{0}^{k+1}) =
j_{k+1}([\gamma]_{0}^{k+1})(f)\; .$$

\section{Variational calculus with higher-order constraints}\label{section3}

In this section, we  briefly review the main notions of variational calculus with higher-order constraints.

Let us consider a mechanical system whose dynamic is described by a Lagrangian
$L: T^{(k)}Q\rightarrow \R$ that depends on higher-order derivatives up to order $k$.   Given two points $x,y \in T^{(k-1)}Q$ we define
the infinite-dimensional  manifold ${\mathcal
C}^{2k}(x,y)$ of  $2k$-differentiable
curves which connect $x$ and $y$ as
\[
{\mathcal  C}^{2k}(x,y) = \{ c:[0,T] \longrightarrow Q \; \big| \; c
\; \hbox{is} \; C^{2k}, c^{(k-1)}(0)=x \; \hbox{and} \;
c^{(k-1)}(T)=y \}\, .
\]

Fixed a  curve $c$ in ${\mathcal C}^{2k}(x,y)$,
the tangent space to ${\mathcal C}^{2k}(x,y)$ at
$c$ is given by
\begin{eqnarray*}
T_{c}{\mathcal C}^{2k}(x,y) &=& \left\{ X : [0,T]
\longrightarrow TQ \; \big| \; X \;
\hbox{is} \; C^{2k-1}, X(t) \in T_{c(t)}Q, \right. \\
&& X^{(k-1)}(0)=0 \; \hbox{and} \; X^{(k-1)}(T)=0
\}\; .
\end{eqnarray*}

Let us consider the \textit{action functional} ${\mathcal A}$ on
$C^{2k}$-curves in $Q$ given by
\begin{equation}\label{qqq}
\begin{array}{rrcl}
{\mathcal A}:& C^{2k}(x, y)&\longrightarrow& \R\\
        & c &\longmapsto & \int^T_0  L( c^{(k)}(t))\; dt\; .
\end{array}
\end{equation}

\begin{definition}{\bf Hamilton's principle.}
A curve $c\in {\mathcal  C}^{2k}(x,y)$ is a
solution of the Lagrangian system determined by
$L: T^{(k)}Q\longrightarrow \R$ if and only if
$c$ is a critical point of ${\mathcal A}$.
\end{definition}

In order to find the critical points of ${\mathcal A}$, we need to
characterize the curves $c$ such that $d{\mathcal
A}(c)(X)=0$ for all $X\in T_{c}{\mathcal
C}^{2k}(x,y)$. Taking a family of curves
$c_{\epsilon}\in {\mathcal C}^{2k}(x, y)$ with
$c_0=c$ and $\epsilon \in (-b, b)\subset \R$,
the stationary condition can be written as
\[
\frac{d}{d\epsilon}{\Big{|}}_{\epsilon=0}
{\mathcal A}(c_{\epsilon})=0\, .
\]

Let us denote
 $\displaystyle{\delta
c^i=\frac{d}{d\epsilon}\Big|_{\epsilon=0}c^i_{\epsilon}}$ and
$\displaystyle{\delta^{(l)} c^i=\frac{d^l}{dt^l}\delta c^i}$; then
we deduce the following result (see \cite{MaPaSh} for the same
result when $k=1$).

\begin{theorem}

Let $L: T^{(k)}Q\rightarrow \R$ be a Lagrangian of order $k$ and ${\mathcal A}(c) = \int^T_0  L( c^{(k)}(t)) dt$ the action functional defined on ${\mathcal C}^{2k} (x,y).$
Then, there exists a unique operator
$
{\mathcal E}L: T^{(2k)} Q\longrightarrow T^*Q
$
and a unique 1-form  $\Theta_L$ on
$T^{(2k-1)} Q$ such that for all variations
$\delta c_{\epsilon}\in T_c{\mathcal
C}^{2k}(x,y)$ we have
\[
d{\mathcal A}(c)\cdot\delta c_{\epsilon}=\int^T_0
{\mathcal E} L (c^{(2k)}(t))\cdot\delta c(t)\;
dt+
\left[\Theta_L(c^{(2k-1)}(t))\cdot\delta^{(2k-1)}
c(t)\right]_0^T\; .
\]
${\mathcal E} L$ is called \textbf{Euler-Lagrange operator} and $\Theta_L$ is called the \textbf{Poincar\'e-Cartan  1-form.}
In local coordinates we
have that
\begin{eqnarray*}
{\mathcal
E}L&=&\sum_{l=0}^{k}(-1)^l\frac{d^l}{dt^l}\left(\frac{\partial
L}{\partial q^{(l)i}}\right)\, dq^i,\\
\Theta_L & = & \sum^{k-1}_{l=0}\hat{p}_{(l)i}\, dq^{(l)i} \; ,
\end{eqnarray*}
where the functions $\hat{p}_{(l)i}$ with $0 \leq l \leq k-1$, are
the \textbf{Jacobi-Ostrogradski generalized momenta} defined by
\[
\hat{p}_{l(i)} = \sum^{k-l-1}_{s=0}(-1)^l {d_T^s}
\left(\frac{\partial L}{\partial
q^{(l+s+1)i}}\right) \; .
\]

The equations of motion of the system, called \textbf{higher-order Euler-Lagrange equations}, are locally written as
\begin{eqnarray*}
\sum_{l=0}^{k}(-1)^l\frac{d^l}{dt^l}\left(\frac{\partial
L}{\partial q^{(l)i}}\right)&=& 0 \; , \quad 1\leq i\leq n.
\end{eqnarray*}
\end{theorem}

The 1-form $\Theta_{L}$ give rise
the exact 2-form  $\Omega_{L} = - d\Theta_{L}$ on $T^{(2k-1)}Q$
which is called the \textbf{Poincar\'e Cartan 2-form}. In local
coordinates
 $$\Omega_L =  \sum^{k-1}_{l=0}dq^{(l)i} \wedge
d\hat{p}_{(l)i}.$$
Is easy to prove that $\Omega_{L}$ is symplectic if and only if
 $$\det
\left(\frac{\partial^2 L}{\partial q^{(k)i}
\partial q^{(k)j}}\right)\neq 0.
$$
The higher-order Lagrangian $L$ is called \textbf{regular} if the 2-form
$\Omega_L$ is symplectic. In the following, we will assume that the Lagrangian $L$ is regular.
%

 We  will see that introducing higher-order constraints and applying the Lagrangian multipliers Lemma (see, for instance, \cite{AbMa}), it is possible to derive the equations of motion and the corresponding geometric structures associated to this constrained problem.

Let us consider a submanifold ${\mathcal M}$ of $T^{(k)}Q$ locally determined by the vanishing of the constraints functions
$\Phi^{\alpha}: T^{(k)}Q\rightarrow\R$, $1\leq \alpha\leq m$.

We assume that the restriction  of the projection $(\tau_Q^{(k-1, k)})_{|{\mathcal M}}: {\mathcal M}\to T^{(k-1)}Q$ is a submersion. Locally, this conditions means that  the $m\times n$-matrix
$$\left(\frac{\partial(\Phi^{1},...,\Phi^{m})}{\partial(q_{1}^{(k)},...,q_{n}^{(k)})}\right)$$
 is of rank $m$ at all points of ${\mathcal M}$.

Consider now the subset ${\mathcal C}^{2k}(x,y, {\mathcal M})$ of ${\mathcal C}^{2k}(x,y)$ of curves that satisfies these constraint equations, that is
\begin{eqnarray*}
{\mathcal C}^{2k}(x,y, {\mathcal M})&=& \{ c:[0,T] \longrightarrow Q \; \big| \; q
\; \hbox{is}
\; C^{2k}, c^{(k-1)}(0)=x,\\
&& \; \hbox{and} \; c^{(k-1)}(T)=y \hbox{ and } c^{(k)}(t)\in
{\mathcal M} \hbox{ for all} \, t\in [0, T]\}.
\end{eqnarray*}

\begin{definition}
A curve $c\in {\mathcal C}^{2k}(x,y, {\mathcal M})$ will be called a solution of the higher-order variational problem with constraints if $c$ is a critical point of
${\mathcal A}{\Big{|}}_{{{\mathcal C}^{2k}}{(x,y, {\mathcal M})}}$.
\end{definition}

For solving this type of problems we will need the following version of the  Lagrange Multipliers Lemma.

\begin{lemma} Let ${\mathcal N}$ be a smooth manifold and let ${\mathcal F}$ be a Banach space with $g: {\mathcal N}\rightarrow {\mathcal F}$ a smooth submersion so that $g^{-1}(0)$ is a submanifold of ${\mathcal N}$. Let $f: {\mathcal N}\rightarrow \R$ be a smooth function. Then $x\in g^{-1}(0)$  is critical point of $f{\Big{|}}_{g^{-1}(0)}$ if and only if there exists $\lambda\in {\mathcal F}^{*}$ such that $x$ is a critical point of $f - \lambda\circ g$.
\end{lemma}

As in the case of systems with constraints on $TQ$,  by using the Lagrange Multipliers lemma, we may characterize the regular critical points of the higher-order problem with constraints as an unconstrained problem for an extended Lagrangian system. (See \cite{LeMu}, for a detailed proof).

\begin{proposition}(\textbf{Variational problem with higher-order constraints})

A curve $c\in {\mathcal C}^{2k}(x,y, {\mathcal M})$ is a critical point of the
variational problem with higher-order constraints if and only if $c$
is a critical point of the functional \[
\int_0^1
\mathcal{L}(q^{(k)}(t), \lambda(t))\; dt\; ,\
\]
where $\lambda=(\lambda_1, \ldots ,\lambda_m)$ as regarded as generalized coordinates on $\R^m$
and ${\mathcal L}: T^{(k)}Q\times \R^m\to \R$ is defined by
\[
{\mathcal L}(q^{(0)}, q^{(1)}, \ldots, q^{(k)}, \lambda)=L(q^{(0)}, q^{(1)}, \ldots, q^{(k)})-\lambda_\alpha \Phi^\alpha(q^{(0)}, q^{(1)}, \ldots, q^{(k)})\; .
\]
\end{proposition}

\begin{remarkth}
The equations
\begin{eqnarray*}
&&\sum_{l=0}^{k}(-1)^l\frac{d^l}{dt^l}\left(\frac{\partial
L}{\partial q^{(l)i}}-\lambda_{\alpha}\frac{\partial
\Phi^{\alpha}}{\partial q^{(l)i}} \right)=0\ \ \ \ i = 1,\ldots,n,\qquad \alpha=1, \ldots, m\, ,\\
&&\Phi(q^{(0)}, q^{(1)}, \ldots, q^{(k)}) = 0\; ,
\end{eqnarray*}
 are called \textbf{Euler-Lagrange equations with higher-order constraints}.
\end{remarkth}

\section{Geometric formulation for higher-order constrained variational problems}\label{section4}

Now, we  develop a geometric characterization higher-order constrained variational problems using, as an essential tool, the Skinner and Rusk formulation  (see \cite{Skinner-Rusk}).

Let us consider  the Whitney sum $T^*(T^{(k-1)}Q)\oplus T^{(k)}Q$
and the canonical projections $$pr_1: T^*(T^{(k-1)}Q)\oplus
T^{(k)}Q\longrightarrow T^*(T^{(k-1)}Q),$$ $$pr_2:
T^*(T^{(k-1)}Q)\oplus T^{(k)}Q\longrightarrow T^{(k)}Q.$$

Let us take the submanifold $W_0=pr_2^{-1}({\mathcal M})=T^*(T^{(k-1)}Q)\times {\mathcal M}$ and the restrictions to $W_0$ of the canonical projections $pr_1$ and $pr_2$

$$\pi_1={pr_1}_{\big|{W_0}}:W_0\subset T^{*}(T^{k-1}Q)\oplus T^{(k)}Q\rightarrow T^{*}(T^{(k-1)}Q)$$  $$\pi_2={pr_2}_{\big|{W_0}}:W_0\subset T^{*}(T^{k-1}Q)\oplus T^{(k)}Q\rightarrow {\mathcal M}\; .$$ \


Now, we consider on $W_0$ the presymplectic 2-form $$\Omega_{W_0}=\pi_1^*(\omega_{T^{(k-1)}Q}),$$ where
$\omega_{T^{(k-1)}Q}$ is the canonical symplectic form on
$T^*(T^{(k-1)}Q)$. Define also the function $H_{W_0}: W_0\rightarrow\R$ given by
\[
H_{W_0}(\alpha,p)=\langle \alpha, j_k(p)\rangle - {L|_{{\mathcal M}}}(p)
\]
where $(\alpha,p)\in W_0=T^*(T^{(k-1)}Q)\times {\mathcal M}$.
Here $\langle \cdot , \cdot \rangle$ denotes the natural paring
between vectors and covectors on $T^{(k-1)}Q$ (observe that $j_k(p)\in T T^{(k-1)}Q$).

 We will see that the dynamics of the higher-order constrained variational problem is intrinsically characterized as the solutions of the presymplectic hamiltonian equation \begin{equation}\label{eq1}
i_X\Omega_{W_0}=dH_{W_0} \, .
\end{equation}

Let us consider $\Omega={pr_1}^*(\omega_{T^{(k-1)}Q})$ and $H: T^{*}(T^{(k-1)}Q)\oplus T^{(k)}Q\rightarrow\R$ given by
$$H=\langle pr_1, pr_2\rangle -pr_2^* L = \langle pr_1, pr_2\rangle - L\circ\pi_2.$$ Observe that locally
\[
\ker \Omega=\hbox{span }\left\langle {\mathcal V}_{i}=\frac{\partial}{\partial q^{(k)i}}\right\rangle\, .
\]

Then, it is easy to show that equations \eqref{eq1} are  equivalent to  (see \cite{BEMMR})
\begin{equation}\label{eq2}
\left\{
\begin{array}{rcl}
i_X\Omega-dH&\in& (TW_0)^0\\
X&\in& TW_0\, ,
\end{array}
\right.
\end{equation}
where $T W^0$ is the annihilator of $T W_0$ locally spanned by $\{d\Phi^\alpha\}$, where $\Phi^\alpha: W_0\to \R$ denote the constraints
$\Phi^{\alpha}=\Phi^{\alpha}\circ pr_2$ (for notational simplicity, we do not distinguish the notation between constraints on ${\mathcal M}$ and constraints on $W_0$).

Take coordinates $(q^{(0)i}, q^{(1)i}, \ldots, q^{(k-1)i};
p^{(0)}_i, \ldots, p^{(k-1)}_i, q^{(k)i})$ in $T^*(T^{(k-1)}Q)\oplus
T^{(k)}Q$, then the local expressions of the presymplectic 2-form $\Omega$ and the hamiltonian $H$ are 
\begin{eqnarray*}
\Omega&=& \sum_{r=0}^{k-1}d q^{(r)i}\wedge d p^{(r)}_i\; ,\\
H&=& \sum_{r=0}^{k-1}q^{(r+1)i}p^{(r)}_i-L(q^{(0)i}, q^{(1)i}\; .
\ldots, q^{(k) i})\; .
\end{eqnarray*}

Consider  a vector field $X$ on $T^{*}(T^{(k-1)}Q)\oplus
T^{(k)}Q$ with local expression
\[
X=\sum_{r=0}^k X^{(r)i}\frac{\partial}{\partial
q^{(r)i}}+\sum_{r=0}^{k-1} Y^{(r)}_i\frac{\partial}{\partial
p^{(r)}_i}.
\]

The equations \eqref{eq2} implies that
\begin{eqnarray*}
-Y_{i}^{(0)} &=& - \frac{\partial L}{\partial q^{(0)i}} + \lambda_{\alpha}\frac{\partial\Phi^{\alpha}}{\partial q^{(0)i}}
\ , ,\\
-Y_{i}^{(r)} &=& p_{i}^{(r-1)} - \frac{\partial L}{\partial q^{(r)i}} + \lambda_{\alpha}\frac{\partial\Phi^{\alpha}}
{\partial q^{(r)i}}\, ,   \quad r = 1, \ldots, k-1\, ,\ \\
X_{i}^{(r)} &=& \frac{\partial H}{\partial p_{i}^{(r)}} = q^{(r+1)i}; \ \ r= 0,\ldots,k-1\; .
\end{eqnarray*}

The solutions of Equation \eqref{eq2} are defined on the first constraint submanifold given by the set of points $x\in W_0$ such that
  $(dH+\lambda_\alpha d\Phi^\alpha)(x)(Z)=0,$ for all $Z\in \ker \Omega(x)$. Locally these restrictions are defined from the following relations %
$$\varphi_{i}^{1}
= p_{i}^{(k-1)} - \frac{\partial L}{\partial q^{(k)i}} +
\lambda_{\alpha}\frac{\partial\Phi^{\alpha}}{\partial q^{(k)i}} =
0, \quad i= 1,\ldots,n\; .$$
The equations $\varphi_{i}^{1} = 0$ (primary relations) determine the set of points $W_1$ of $W_0$ where \eqref{eq2} has a solution. $W_1$ is the primary constraint submanifold (assuming that it is a submanifold) for the presymplectic hamiltonian system $(W_0, \Omega_{W_0}, H_{W_0})$. (See, for instance, \cite{GN1}).

Then, we have two different types of equations which restrict  the dynamics on $T^*(T^{(k-1)}Q)\oplus T^{(k)}Q$
\begin{eqnarray}
\Phi^{\alpha} &=& 0  \quad \alpha=1, \ldots, m  \quad \hbox{  (constraints determining ${\mathcal M}$)  } \label{qwe1} \\
\varphi^1_i &=& 0 \quad i=1, \ldots, n . \quad \hbox{  (primary relations)  }\label{qew2}
\end{eqnarray}

Therefore, the equations that define an integral curve of the solution $X$ are given by
\begin{eqnarray}\label{eq3}
\frac{d q^{(r)i}}{dt} &=& q^{(r+1)i}, \quad \quad r=0, \ldots, k-1,\nonumber\\
\frac{d p_{i}^{(0)}}{dt} &=& \frac{\partial L}{\partial q^{(0)i}} - \lambda_{\alpha}\frac{\partial\Phi^{\alpha}}{\partial q^{(0)i}}\; ,\nonumber\\
\frac{d p_{i}^{(r)}}{dt} &=& \frac{\partial L}{\partial q^{(r)i}} - \lambda_{\alpha}\frac{\partial\Phi^{\alpha}}{\partial q^{(r)i}}-p_{i}^{(r-1)}\;, \quad \quad r=1,\ldots, k-1\; ,
\end{eqnarray}
and the constraint equations \eqref{qwe1} and \eqref{qew2}.

Differentiating with respect to time the equations $\varphi^1_i$, substituting in \eqref{eq3} and proceeding further, we find the equations of motion for the higher-order variational problem analyzed in Section \ref{section3}, i.e.
 \begin{equation}
\sum_{r=0}^{k}(-1)^r\frac{d^r}{dt^r}\left(\frac{\partial
L}{\partial q^{(r)i}}-\lambda_{\alpha}\frac{ \partial
\Phi^{\alpha}}{\partial q^{(r)i}}\right)=0\; .
\end{equation}

 The solution of equation \eqref{eq1} on $W_1$ may not be tangent to $W_1$. In such a case, we have to restrict $W_1$ to the submanifold $W_2$ where there exists at least a  solution  tangent to $W_1$. Proceeding further, we obtain a sequence of submanifolds \cite{GN1} (assuming that all the subsets generated by the algorithm are submanifolds)
\[
\cdots\hookrightarrow W_k\hookrightarrow \cdots \hookrightarrow
W_2\hookrightarrow W_1\hookrightarrow W_0\; .
\]
Algebraically, these constraint submanifolds can be described as
\begin{equation}
W_i=\left\{x\in T^*\left(T^{(k-1)}Q\right)\times_{T^{(k-1)} Q} {\mathcal M}\;\;
\big| \; \; dH_{W_0}(x)(v)=0\;  \ \forall v\in \left(T_x W_{i-1}\right)^{\perp} \;
\right\}\; \quad i\geq 1\; ,
\end{equation}
 where $\left(T_x W_{i-1}\right)^{\perp} = \left\{ v \in T_xW_0 \; \big| \;
\Omega_{W_0}(x)(u,v) = 0 \; \ \forall \, u \in T_x W_{i-1} \;
\right\}$.

If  this constraint algorithm stabilizes, i.e., there exists a positive integer $k\in \N$ such that $W_{k+1} = W_{k}$ and $\dim W_{k}\geq 1,$ then we will have at least a well defined solution $X$ on $W_{f} = W_{k}$ such that \[
\left(i_X\Omega_{W_0}=dH_{W_0}\right)_{| W_f}\; .
\]

Now, denote by $\Omega_{W_1}$, the pullback of  presymplectic 2-form $\Omega_{W_0}$ to $W_1$.
In order to establish a necessary and sufficient condition for the symplecticity of the 2-form $\Omega_{W_1}$, we define as in Section \ref{section3}, the extended lagrangian
$${\mathcal L} = L - \lambda_{\alpha}\Phi^{\alpha}\; .$$

\begin{theorem}\label{theorem-1}
For any choice of coordinates $(q^{(0)i}, q^{(1)i}, \ldots, q^{(k-1)i};
p^{(0)}_i, \ldots, p^{(k-1)}_i, q^{(k)i})$ in
$T^*(T^{(k-1)}Q)\oplus T^{(k)}Q$, we have that $(W_1, \Omega_{W_1})$ is a symplectic manifold if and only if \begin{equation}
\det\; \left(
\begin{array}{cc}
\frac{{\partial}^{2}{\mathcal L}}{\partial q^{(k)i}\partial q^{(k)j}}& -\frac{\partial \Phi^\alpha}{\partial q^{(k)i}}\\
\frac{\partial \Phi^\beta}{\partial q^{(k)j}}&\mathbf{0}\end{array}
\right)=\hbox{det}\; \left(\begin{array}{cc}
\frac{{\partial}^{2}{L}}{\partial q^{(k)i}\partial q^{(k)j}}-\lambda_{\alpha}\frac{\partial \Phi^{\alpha}}{\partial q^{(k)i}\partial q^{(k)j}}& -\frac{\partial \Phi^\alpha}{\partial q^{(k)i}}\\
\frac{\partial \Phi^\beta}{\partial q^{(k)j}}&\mathbf{0}\end{array}
\right)\not=0
\end{equation}
\end{theorem}
(The proof follows the same lines that the one   in Proposition \ref{proposition}).

\begin{remarkth}
{\rm
Observe that if the determinant of the matrix in  Theorem \ref{theorem-1} is not zero, then we can apply the implicit function theorem to the equation of constraints $\varphi_{i}^{1} = 0$ and $\Phi^{\alpha}=0$, and we can express the Lagrange multipliers $\lambda_{\alpha}$ and higher-order velocities $q^{(k)i}$ in terms of coordinates $(q^{(0)i},\ldots, q^{(k-1)i},  p^{(0)}_i,
\ldots, p^{(k-1)}_i)$, i.e.,
\begin{eqnarray*}
\lambda_{\alpha} &=& \lambda_{\alpha}(q^{(0)}, q^{(1)}, \ldots, q^{(k-1)}, p^{(0)}, \ldots, p^{(k-1)})\; ,\\
q^{(k)i} &=& q^{(k)i}(q^{(0)}, q^{(1)}, \ldots, q^{(k-1) }, p^{(0)}, \ldots, p^{(k-1)})\; .
\end{eqnarray*}

Thus we can consider
  $(q^{(0)i}, q^{(1)i}, \ldots, q^{(k-1) i},
p^{(0)}_i, \ldots, p^{(k-1)}_i)$ as local coordinates in $W_1$.
In this case,
\[
\Omega_{W_1} = \sum_{r=0}^{k-1}d q^{(r)i}\wedge d p^{(r)}_i
\]
which is obviously symplectic.
}
\end{remarkth}


\section{Optimal Control of underactuated mechanical systems}\label{section5}

After introducing the geometry of higher-order Lagrangian system with constraints  in the previous section, we may turn to  the geometric framework for optimal control of underactuated mechanical systems.
We recall that a Lagrangian control system is underactuated if the
number of the control inputs is less than the dimension of the
configuration space.
We  assume, in the sequel, that the considered systems are controllable \cite{bullolewis}.

 Consider the class of  underactuated Lagrangian control system (superarticulated  mechanical system following the nomenclature by \cite{Ba}) where the  configuration space $Q$ is the cartesian product of two differentiable manifolds, $Q = Q_1\times Q_2$. Denote by $(q^A)=(q^a,
q^{\alpha})$, $1\leq A\leq n$,  local  coordinates on $Q$
where $(q^a)$, $1\leq a\leq r$ and $(q^{\alpha})$, $r+1\leq
\alpha\leq n$, are local coordinates on $Q_1$ and $Q_2$,
respectively.

Given a Lagrangian $L:TQ \equiv TQ_1\times TQ_2 \rightarrow
\mathbb{R}$, we assume that the controlled external forces can be
applied only to the coordinates $(q^a)$. Thus, the equations of motion
 are given by
  \begin{equation}\label{lagrange con control}
\begin{split}
&\frac{d}{dt}\left(\frac{\partial L}{\partial \dot q^a}\right)-
\frac{\partial L}
{\partial q^a}=u^a, \\
&\frac{d}{dt}\left(\frac{\partial L}{\partial \dot
q^\alpha}\right)- \frac{\partial L}{\partial q^\alpha}=0\, ,
\end{split}
\end{equation}
where $a = 1,\ldots,r,$ and $ \alpha =
r+1,\ldots,n.$

We study the optimal control problem that consists on finding a
trajectory $(q^a(t), q^{\alpha}(t), u^a(t))$ of state variables
and control inputs satisfying equations \eqref{lagrange con control} from
given initial and final conditions, $(q^{a}(t_0), q^{\alpha}(t_0),
\dot{q}^{a}(t_0), \dot{q}^{\alpha}(t_0)),$ $(q^{a}(t_f),
q^{\alpha}(t_f), \dot{q}^{a}(t_f), \dot{q}^{\alpha}(t_f))$
respectively, minimizing the cost functional
\[
{\mathcal A}=\int_{t_0}^{t_f} C(q^a, q^{\alpha}, \dot{q}^a,
\dot{q}^{\alpha}, u^a)\, dt.
\]

It is well know
(see \cite{Blo}) that
this optimal control problem is equivalent to the following
constrained variational problem.

Extremize
\[ \widetilde{\mathcal A}=\int_{t_0}^{t_f}
\widetilde{L}(q^a(t), q^{\alpha}(t), \dot{q}^a(t),
\dot{q}^{\alpha}(t), \ddot{q}^a(t), \ddot{q}^{\alpha}(t))\, dt
\]
subject to the second order constraints given by
\[
\Phi^{\alpha}(q^a, q^{\alpha}, \dot{q}^a, \dot{q}^{\alpha},
\ddot{q}^a, \ddot{q}^{\alpha})=\frac{d}{dt}\left(\frac{\partial
L}{\partial \dot q^{\alpha}}\right)- \frac{\partial L}{\partial
q^{\alpha}}=0\; ,
\]
and the boundary conditions,
where $\widetilde{L}:T^{(2)}Q\rightarrow\mathbb{R}$ is defined as
\[
\widetilde{L}(q^a, q^{\alpha}, \dot{q}^a, \dot{q}^{\alpha},
\ddot{q}^a, \ddot{q}^{\alpha})= C\left(q^a, q^{\alpha}, \dot{q}^a,
\dot{q}^{\alpha}, \frac{d}{dt}\left(\frac{\partial L}{\partial
\dot q^a}\right)- \frac{\partial L}{\partial q^a}\right).
 \]

 Now, according to the formulation given in Section \ref{section4}, the dynamics of this second order constrained
variational problem is determined by the solution of a
pre\-symplectic Hamiltonian system. In the following we  repeat some of the constructions given in \ref{section4} but specialized to this particular setting, obtaining new insights for the optimal control problem under study.

If ${\mathcal M}\subset T^{(2)}Q$ is the submanifold given by annihilation of
the functions $\Phi^{\alpha}$, we will see how to define local
coordinates on ${\mathcal M}$.

{}From the constraint equations we have
\[\frac{d}{dt}\left(\frac{\partial L}{\partial\dot q^{\alpha}}\right)-\frac{\partial L}{\partial q^{\alpha}} =
0 \Longleftrightarrow \frac{\partial^{2}L}{\partial\dot
q^{\beta}\partial\dot q^{\alpha}}\ddot q^{\beta} =
F_{\alpha}(q^{i},\dot q^{i},\ddot q^{a}).\]

Let us assume that the matrix $(W_{\alpha\beta}) =
\left(\frac{\partial^{2}L}{\partial \dot q^{\alpha}\partial \dot
q^{\beta}}\right)$ is non-singular and denote by $(W^{\alpha\beta})$
its inverse.
Thus,
$$\ddot q^{\alpha} = W^{\alpha\beta}F_{\alpha}(q^{i}, \dot q^{i}, \ddot q^{a} )=G^{\alpha}(q^{i}, \dot q^{i}, \ddot q^{a}).$$
Therefore, we can consider $(q^{i}, \dot q^{i}, \ddot q^{a} )$ as a
system of local coordinates on ${\mathcal M}$. The canonical inclusion $i_{\mathcal M}:
{\mathcal M}\hookrightarrow TTQ$  can be written as
\[
\begin{array}{rcl}
  {\mathcal M}&\stackrel{i_{\mathcal M}}{\rightarrow}& TTQ\\
      (q^{i}, \dot q^{i}, \ddot q^{a} )& \mapsto&(q^{i}, \dot q^{i}, \ddot q^{a}, G^{\alpha}(q^{i}, \dot q^{i}, \ddot q^{a}))\; .
\end{array}
\]
Define the restricted lagrangian  $\widetilde{L}\mid_{{\mathcal M}} :{\mathcal M}\rightarrow\mathbb{R}$.

\begin{figure}[h]
$$\xymatrix{
  &&W_0=  T^{*}(TQ)\times_{TQ} {\mathcal M} \ar[lld]_{\pi_2} \ar[dd]^{\pi_{W_0,T{\mathcal M}}} \ar[rrd]^{\pi_1}&&\\
  {\mathcal M} \ar[rrd]^{(\tau_{TQ})|_{{\mathcal M}}} && && T^*TQ \ar[lld]^{\pi_{T^*Q}} \\
  && TQ &&
   }$$
   \caption{S¡econd order Skinner and Rusk formalism}
\end{figure}
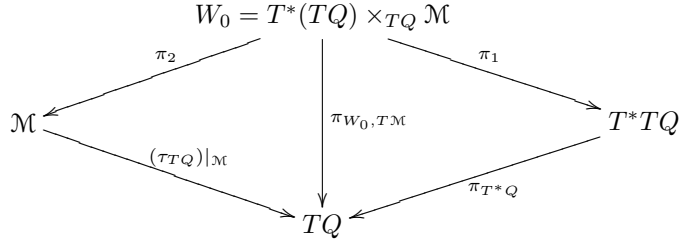

We will consider $W_{0}= T^{*}(TQ)\times_{TQ} {\mathcal M}$ whose
coordinates are $(q^{i}, \dot q^{i};   p_{i}^{0},
p_{i}^{1}, \ddot q^{a})$.

Let us define the 2-form $\Omega_{W_0}=\pi_1^*(\omega_{TQ})$ on $W_{0}$
and ${H}_{W_0}(\alpha_x, v_x)=\langle \alpha_x,
i_{\mathcal M}(v_x)\rangle- \widetilde{L}_{\mathcal M}(v_x)$ where $x\in T Q$, $v_x\in
{\mathcal M}_x=((\tau^{(1,2)}_{Q})|_{\mathcal M})^{-1}(x)$ and $\alpha_x\in T^*_xTQ$. In local
coordinates,
\begin{eqnarray*}
\Omega_{W_0} &=&dq^{i}\wedge dp_{i}^{0} +d\dot q^{i}\wedge dp_{i}^{1},\\
 {H}_{W_0} &=& p_{i}^{0}\dot q^{i} + p_{a}^{1}\ddot q^{a} + p_{\alpha}^{1}G^{\alpha}(q^{i}, \dot q^{i}, \ddot q^{a}) - \widetilde{L}_{{\mathcal M}}(q^{i}, \dot q^{i}, \ddot q^{a}).
 \end{eqnarray*}

The dynamics of this variational constrained problem is determined
by the solution of the equation
\begin{equation}\label{eq4}
i_{X}\Omega_{W_0} =d{H}_{W_0}.
\end{equation}

It is clear that $\Omega_{W_0}$ is a presymplectic form on $W_0$ and locally \[ \ker
\Omega_{W_0}=\hbox{span}\left\langle \frac{\partial}{\partial \ddot
q^a}\right\rangle.
\]

Following the Gotay-Nester-Hinds algorithm we obtain the
primary constraints
\[ d{H}_{W_0}\left(\frac{\partial}{\partial \ddot
q^a}\right)=0\; .
\]
That is,
$$\varphi^1_{a}=\frac{\partial {H}_{W_0}}{\partial \ddot q^{a}}=p_{a}^{1} + p_{\alpha}^{1}\frac{\partial G^{\alpha}}{\partial\ddot q^{a}} - \frac{\partial \widetilde{L}_{{\mathcal M}}}{\partial \ddot q^{a}} = 0.$$

These new constraints $\varphi^1_{a} = 0$ give rise to a submanifold
$W_1$ of dimension $4n$ with local coordinates $(q^{i}, \dot
q^{i}, \ddot q^{a}, p_{i}^{0}, p_{\alpha}^{1})$.

Consider a solution curve $(q^i(t),\dot{q}^i(t), \ddot{q}^a(t), p_{i}^{0}(t), p_{i}^{1}(t))$  of Equation \eqref{eq4}.
 Then, this curve satisfies the following system of differential equations

 \begin{eqnarray}
 \frac{dq^{i}}{dt}&=&\dot{q}^i\; ,\qquad
 \frac{d^2q^{a}}{dt^2}=\ddot{q}^a\; ,\\
 \frac{d^2q^{\alpha}}{dt^2}&=& G^{\alpha}(q^i, \frac{d q^i}{dt},  \frac{d^2 q^a}{dt^2})\; ,\label{qe-1}\\
 \frac{dp_i^0}{dt}&=&-p^1_{\alpha}\frac{\partial G^{\alpha}}{\partial q^i}+ \frac{\partial \widetilde{L}_{\mathcal M}}{\partial q^i}\; ,\label{qe-2}\\
 \frac{d p_{i}^1}{dt}&=&-p_i^0-p^1_{\alpha}\frac{\partial G^{\alpha}}{\partial \dot{q}^i}+ \frac{\partial \widetilde{L}_{\mathcal M}}{\partial \dot{q}^i}\; ,\label{qe-3}\\
p_{a}^{1} &=&-p_{\alpha}^{1}\frac{\partial G^{\alpha}}{\partial\ddot q^{a}} + \frac{\partial \widetilde{L}_{{\mathcal M}}}{\partial \ddot q^{a}}\; .\label{qe-4}
\end{eqnarray}
{}From Equations \eqref{qe-3} and \eqref{qe-4} we deduce
\[
\frac{d}{dt}\left(\frac{\partial \widetilde{L}_{{\mathcal M}}}{\partial \ddot q^{a}}-p_{\alpha}^{1}\frac{\partial G^{\alpha}}{\partial\ddot q^{a}}\right)
=-p_a^0-p^1_{\alpha}\frac{\partial G^{\alpha}}{\partial \dot{q}^a}+ \frac{\partial \widetilde{L}_{\mathcal M}}{\partial \dot{q}^a}\; .
\]
Differentiating with respect to time, replacing in the previous equality and using \eqref{qe-2} we obtain the following system of $4$-order differential equations
\begin{eqnarray}
&&\frac{d^2}{dt^2}\left(\frac{\partial \widetilde{L}_{{\mathcal M}}}{\partial \ddot q^{a}}-p_{\alpha}^{1}\frac{\partial G^{\alpha}}{\partial\ddot q^{a}}\right)
-\frac{d}{dt}\left(\frac{\partial \widetilde{L}_{\mathcal M}}{\partial \dot{q}^a}-p^1_{\alpha}\frac{\partial G^{\alpha}}{\partial \dot{q}^a}\right)+
 \frac{\partial \widetilde{L}_{\mathcal M}}{\partial q^a}-p^1_{\alpha}\frac{\partial G^{\alpha}}{\partial q^a}=0\; .\label{poi-1}
 \end{eqnarray}
 Also, using (\ref{qe-2}) and (\ref{qe-3}) we deduce
 \begin{equation}
   \frac{d^2 p_{\alpha}^1}{dt^2}=
    \frac{d}{dt}\left(\frac{\partial \widetilde{L}_{\mathcal M}}{\partial \dot{q}^\alpha}-p_\beta^1\frac{\partial G^\beta}{\partial \dot q^{\alpha}}\right)
   -\left(\frac{\partial \widetilde{L}_{\mathcal M}}{\partial {q}^\alpha}-p_\beta^1\frac{\partial G^\beta}{\partial q^{\alpha}}\right)\; .
      \label{poi-2}
\end{equation}
If we solve the implicit system of differential equations  given by  (\ref{poi-1}) and (\ref{poi-2}) then from Equations  (\ref{qe-3}) and (\ref{qe-4}) we deduce that the values of $p^0_a$ and $p^0_\alpha$ are
\begin{eqnarray}
p^0_a&=& \frac{\partial \widetilde{L}_{\mathcal M}}{\partial \dot{q}^a}-p^1_{\alpha}\frac{\partial G^{\alpha}}{\partial \dot{q}^a}
-\frac{d}{dt}\left( \frac{\partial \widetilde{L}_{{\mathcal M}}}{\partial \ddot q^{a}} -p_{\alpha}^{1}\frac{\partial G^{\alpha}}{\partial\ddot q^{a}} \right)\; ,\label{p0-a}\\
p^0_{\alpha}&=&\frac{\partial \widetilde{L}_{\mathcal M}}{\partial \dot{q}^\alpha}-p^1_{\beta}\frac{\partial G^{\beta}}{\partial \dot{q}^\alpha}
-\frac{d p^1_\alpha}{dt}\label{po-al}\; .
\end{eqnarray}

Since, from our initial problem, we are only interested in the values $q^A(t)$, it is uniquely necessary to solve the coupled system of implicit    differential equations given by  (\ref{poi-1}), (\ref{poi-2}) and (\ref{qe-1}) without explicitly calculate the values $p^0_a(t)$.

Now, we are interested in the geometric properties of the dynamics. First, consider
the submanifold $W_1$ of $W_0$ determined by
\[ W_1=\{ x\in T^*TQ\times_{TQ} {\mathcal M}\; \big| \; d {H}_{W_0}(x)(V)=0 \
\forall \ V\in \ker \Omega(x)\}  \]
and the 2-form $\Omega_{W_1}=i^*_{W_1}\Omega_{W_0}$, where  $i_{W_1}: W_1\hookrightarrow W_0$ denotes the canonical inclusion.
Locally, $W_1$ is determined by the vanishing of the constraint equations
$$ \varphi^1_{a} = p_{a}^{1} + p_{\alpha}^{1}\frac{\partial G^{\alpha}}{\partial\ddot q^{a}} - \frac{\partial \widetilde{L}_{{\mathcal M}}}{\partial \ddot q^{a}} = 0\; .$$

Therefore, we can consider local coordinates $(q^i, \dot{q}^i, \ddot{q}^a, p^0_i, p^1_\alpha)$ on $W_1$.

\begin{proposition}\label{proposition}
$(W_{1}, \Omega_{W_{1}})$ is symplectic if and only if for any
choice of local coordinates $(q^{i}, \dot q^{i}, \ddot q^{a},
p_{i}^{0}, p_{i}^{1})$ on $W_0$,
\begin{equation}\label{prop}
\det\left( \mathcal{R}_{ab}\right)=\det\left(\frac{\partial^{2} \widetilde{L}_{{\mathcal M}}}{\partial \ddot
q^{a}\partial \ddot q^{b}} - p_{\alpha}^{1}\frac{\partial^{2}
G^{\alpha}}{\partial \ddot q^{a}\partial\ddot q^{b}}
\right)_{(n-r)\times (n-r)} \neq 0\; \hbox{   along  }  W_1\; .
\end{equation}
\end{proposition}

\textbf{Proof:}

Let us recall that $\Omega_{W_1}$ is symplectic of and only if $T_{x}W_{1}\cap \left(T_{x}W_1\right)^{\perp} = 0 \ \ \forall x\in W_1$,
   where
   \[
   \left(T_{x}W_1\right)^{\perp}=\left\{ v \in T_x(T^{*}TQ) \times_{TQ} {\mathcal M}) \ / \ \Omega_{W_0}(x)(v,w)=0, \hbox{for all } w\in T_x W_1\right\}.
   \]

   Suppose that $(W_1, \Omega_{W_1})$ is symplectic and that
   \[
   \lambda^a {\mathcal R}_{ab}(x)=0 \hbox{  for some  } \lambda^a\in \R \hbox{  and  } x\in W_1\; .
   \]
   Hence
    \[
   \lambda^b {\mathcal R}_{ab}(x)=\lambda^b  d\varphi_a(x) \left(\left.\frac{\partial}{\partial \ddot{q}^b}\right|_{x}\right)=0\; .
   \]
   Therefore, $\lambda^b \left.\frac{\partial}{\partial \ddot{q}^b}\right|_{x}\in T_x W_1$ but it is also in $T_xW_1^\perp$. This implies that $\lambda_b=0$ for all $b$ and that the matrix $({\mathcal R}_{ab})$ is regular.

  Now, suppose that the matrix  $({\mathcal R}_{ab})$ is regular.
Since
  \[
  {\mathcal R}_{ab}(x)=d\varphi_a(x) \left(\left.\frac{\partial}{\partial \ddot{q}^b}\right|_{x}\right)\; ,
  \]
then, $\left.\frac{\partial}{\partial \ddot{q}^b}\right|_{x}\notin T_x W_1$ and, in consequence,
  \[
  T_xW_1\oplus \hbox{span } \left\{ \left.\frac{\partial}{\partial \ddot{q}^b}\right|_{x}\right\}=T_x W_0.
  \]
  Now, let $Z\in T_{x}W_{1}\cap \left(T_{x}W_1\right)^{\perp}$ with $x\in W_1$. It follows that
  \[
  0=i_Z\Omega_{W_0}(x)\left(\left.\frac{\partial}{\partial \ddot{q}^a}\right|_{x}\right), \hbox {  for all  } a   \hbox{ and  } i_Z\Omega_{W_0}(x)(\bar{Z})=0, \hbox{  for all  } \bar{Z}\in T_xW_1\; .
  \]
   Then, $Z\in \ker \Omega_{W_0}(x)$. This implies that
   \[
   Z=\lambda_b \left.\frac{\partial}{\partial \ddot{q}^b}\right|_{x}
   \]
   Since $Z\in T_xW_1$ then
   \[
   0=d \varphi_a(x)(Z)=d \varphi_a(x)\left(\lambda_b \left.\frac{\partial}{\partial \ddot{q}^b}\right|_{x} \right)=\lambda_b{\mathcal R}_{ab}
   \]
    and, consequently, $\lambda_b=0$, for all $b$, and $Z=0$.

\begin{flushright}$\blacksquare$
\end{flushright}

\ In the case where the matrix \eqref{prop} is regular then the equations  (\ref{poi-1}), (\ref{poi-2}) and (\ref{qe-1})  can be written as an explicit system of differential equations of the form
\begin{eqnarray}
\frac{d^4 q^a}{dt^4}&=& \Gamma^a\left( q^i, \frac{d q^i}{dt}, \frac{d^2
q^a}{dt^2}, \frac{d^3 q^a}{dt^2}, p^1_{\alpha},\frac{d p^1_\alpha}{dt}\right)
\label{poi-3}\\
\frac{d^2q^{\alpha}}{dt^2} &=& G^{\alpha}(q^i,
\frac{d q^i }{dt}, \frac{d^2 q^a}{dt^2})\label{po-5}\\
   \frac{d^2 p_{\alpha}^1}{dt^2}&=&
    \frac{d}{dt}\left(\frac{\partial \widetilde{L}_{\mathcal M}}{\partial \dot{q}^\alpha}-p_\beta^1\frac{\partial G^\beta}{\partial \dot q^{\alpha}}\right)
   -\left(\frac{\partial \widetilde{L}_{\mathcal M}}{\partial {q}^\alpha}-p_\beta^1\frac{\partial G^\beta}{\partial q^{\alpha}}\right)\label{po-6}\; .
\end{eqnarray}

\begin{remark}Now, we will analyze an alternative characterization of the condition \eqref{prop} and its relationship with the matrix condition that appears in Theorem \ref{theorem-1}. Using the chain rule

\begin{eqnarray*}
\frac{\partial\widetilde{L}_{{\mathcal M}}}{\partial\ddot q^{a}}& =& \frac{\partial\widetilde{L}}{\partial\ddot q^{a}} + \frac{\partial\widetilde{L}}{\partial\ddot q^{\alpha}}\frac{\partial G^{\alpha}}{\partial\ddot q^{a}}\\
\frac{\partial^{2}\widetilde{L}_{{\mathcal M}}}{\partial\ddot q^{a}\partial\ddot q^{b}} &=& \frac{\partial^{2}\widetilde{L}}{\partial\ddot q^{a}\partial\ddot q^{b}} + \frac{\partial^{2}\widetilde{L}}{\partial\ddot q^{a}\partial\ddot q^{\beta}}\frac{\partial G^{\beta}}{\partial\ddot q^{b}} + \frac{\partial^{2}\widetilde{L}}{\partial\ddot q^{\alpha}\partial\ddot q^{b}}\frac{\partial G^{\alpha}}{\partial\ddot q^{a}} + \frac{\partial^{2}\widetilde{L}}{\partial\ddot q^{\alpha}\partial\ddot q^{\beta}}\frac{\partial G^{\alpha}}{\partial\ddot q^{a}}\frac{\partial G^{\beta}}{\partial\ddot q^{b}} + \frac{\partial^{2}\widetilde{L}}{\partial\ddot q^{\alpha}}\frac{\partial^{2}G^{\alpha}}{\partial\ddot q^{a}\partial \ddot q^{b}}\; .
\end{eqnarray*}

 Define $W_{ij} = \left(\frac{\partial^{2}\widetilde L}{\partial\ddot q^{i}\partial\ddot q^{j}}\right)$, where $\Phi^{\alpha} = \ddot q^{\alpha} - G^{\alpha}.$
\ Then we can write \eqref{prop} as
\[ {\mathcal R}_{ab}=W_{ab} - W_{a\beta}\frac{\partial\Phi^{\beta}}{\partial\ddot q^{b}} - W_{\alpha b}\frac{\partial\Phi^{\alpha}}{\partial\ddot q^{a}}
 +W_{\alpha\beta}\frac{\partial\Phi^{\alpha}}{\partial\ddot q^{a}}\frac{\partial\Phi^{\beta}}{\partial\ddot q^{b}}
 +\left(p^{1}_{\alpha}- \frac{\partial\widetilde{L}}{\partial\dot q^{\alpha}}\right)\frac{\partial^{2}\Phi^{\alpha}}{\partial\ddot q^{a}\partial\ddot q^{b}} \;.\]

Consider now the extended lagrangian ${\mathcal L} = \widetilde{L} - \lambda_{\alpha}\Phi^{\alpha}$ where $\lambda_{\alpha} = \frac{\partial\widetilde{L}}{\partial\dot q^{\alpha}}-p_{\alpha}^{1}$.

Then, the matrix $(\overline{W}_{ij})=\left(\frac{\partial^{2}{\mathcal L}}{\partial\ddot q^{i}\partial\ddot q^{j}}   \right)$ is equal, along $W_1$,  to \begin{equation}\label{leo2} \overline{W}_{ij} = \left(
                         \begin{array}{cc}
                           \overline{W}_{ab} & W_{a\beta} \\
                         W_{\alpha b}  &  W_{\alpha\beta} \\
                         \end{array}
                       \right)\end{equation}
   where $\overline{W}_{ab} = \frac{\partial^{2}\widetilde{L}}{\partial\ddot q^{a}\partial\ddot q^{b}} - \lambda_{\alpha}\frac{\partial^{2}\Phi^{\alpha}}{\partial\ddot q^{a}\partial\ddot q^{b}}.$

 It is easy see that the elements of the matrix \eqref{prop} are given by
\begin{equation}\label{leo1}{\mathcal R}_{ab}=
\overline{W}_{ab} - \overline{W}_{a\beta}\frac{\partial\Phi^{\beta}}{\partial\ddot q^{b}} - \overline{W}_{\alpha b}\frac{\partial\Phi^{\alpha}}{\partial\ddot q^{a}}
 +\overline{W}_{\alpha\beta}\frac{\partial\Phi^{\alpha}}{\partial\ddot q^{a}}\frac{\partial\Phi^{\beta}}{\partial\ddot q^{b}}\;.\end{equation}

Now, using elemental linear algebra it is easy to show that matrix \eqref{leo1} is regular if and only if  the matrix of elements \eqref{leo2} is regular.

\end{remark}

\begin{example}\label{cart-pole}\textbf{ The Cart-Pole System} (see \cite{Blo} and references therein).
{\rm
A Cart-Pole System consists of a cart and an inverted pendulum on it.
The coordinate  $x$ denotes the position of the cart on the $x$-axis and  $\theta$ denotes the angle of the pendulum with the upright vertical.  The configuration space is $Q = \R\times \Su^{1}$.
\begin{center}
\includegraphics[width=10cm]{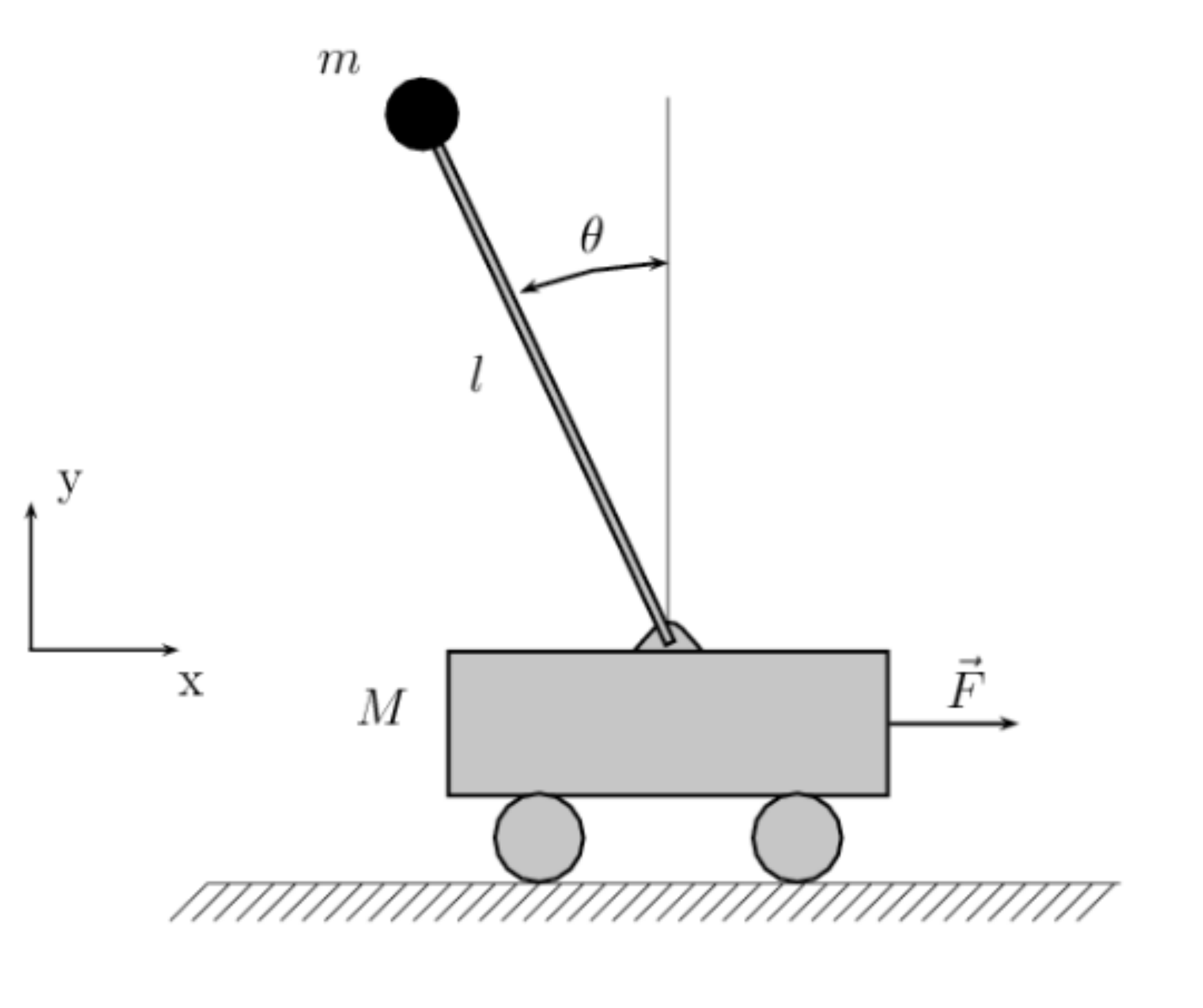}
\end{center}

First, we describe the Lagrangian function describing this system.
The
inertia matrix of the cart-pole system is given by
\begin{eqnarray*}
m_{11} &=& M + m\\
m_{12}(q_2) &=& m_{12}(q_2) = ml\cos(\theta)\\
m_{22} &=& ml^{2}
\end{eqnarray*}
where $M$ is the mass of the cart and $m, l$ are the mass,
and length of the center of mass of pendulum,
respectively. The potential energy of the cart-pole system is
$V(\theta) = mgl\cos(\theta)$.

The
Lagrangian of the system (kinetic energy minus potential energy) is given by
$$L(q ,\dot q) = L(x, \theta,
\dot x, \dot\theta) = \frac{1}{2}M\dot x^{2} + \frac{1}{2}m(\dot
x^{2} + 2\dot x l\dot\theta\cos\theta + l^{2}\dot\theta^{2}) -
mgl\cos\theta-mg\widetilde{h}\; ,$$
where $\widetilde{h}$ is the car height.

The controller can apply a force $F$, the control input,  parallel to the track remaining the
 joint angle $\theta$  unactuated. Therefore, the   equations of motion of the controlled system are
 \begin{eqnarray*}
  (M + m)\ddot x - ml\dot\theta^{2}\sin\theta +
ml\ddot\theta\cos\theta &=& u\\
 \ddot x\cos\theta + l\ddot\theta -g\sin\theta &=& 0
 \end{eqnarray*}


Now we look for  trajectories
$(x(t), \theta(t)), u(t))$
on the state variables and the controls inputs with
initial and final conditions,
$(x(0), \theta(0), \dot x(0), \dot\theta(0)),$
 $(x(T), \theta(T), \dot x(T), \dot\theta(T))$ respectively, and minimizing the cost functional
$$\mathcal{A} = \frac{1}{2}\int_{0}^{T} u^{2} dt.$$

Following our formalism this  optimal control problem
is equivalent to the constrained second-order variational problem determined by
$$\widetilde{\mathcal{A}} = \int_{0}^{T} \widetilde{L}(x, \theta, \dot x, \dot\theta, \ddot x, \ddot\theta )$$
and the second-order constraint
$$\Phi(x, \theta, \dot x, \dot\theta, \ddot x, \ddot\theta ) = \ddot x\cos\theta +
l\ddot\theta -g\sin\theta = 0\; ,$$
where \[\widetilde{L}(x,
\theta, \dot x, \dot\theta, \ddot x, \ddot\theta ) =\frac{1}{2}\left(
\frac{d}{dt}\left(\frac{\partial L}{\partial \dot x}\right)-
\frac{\partial L}{\partial x}\right)^2 = \frac{1}{2}\left[(M +
m)\ddot x - ml\dot\theta^{2}\sin\theta +
ml\ddot\theta\cos\theta\right]^{2}\; .\]

We rewrite the second-order constraint as \begin{eqnarray*}
\displaystyle{\ddot\theta = \frac{g\sin\theta - \ddot
x\cos\theta}{l}}\; .
\end{eqnarray*}

 Thus, the submanifold
  ${\mathcal M}$ of $T^{(2)}(\R\times \Su^1)$  is given by
\[{\mathcal M} = \left\{ (x, \theta, \dot{x},  \dot{\theta}, \ddot{x}, \ddot{\theta})\; \big|\; \ddot x\cos\theta + l\ddot\theta - g\sin\theta = 0\right\}\; .\]
\ Let us consider the submanifold $W_0 =  T^{*}(T(\R\times \Su^{1}))\times_{T(\R\times \Su^{1})}{\mathcal M}$ with
induced coordinates $(x, \theta, \dot x, \dot\theta;
p_{x}^{0},p_{\theta}^{0}, p_{x}^{1}, p_{\theta}^{1}, \ddot x)$.

 Now, we consider the restriction of  $\widetilde{L}$ to ${\mathcal M}$ given
by \[\widetilde{L}|_{{\mathcal M}} = \frac{1}{2}\left[(M + m)\ddot x -
ml\sin\theta\dot\theta^{2} + ml\cos\theta(\frac{g\sin\theta - \ddot
x\cos\theta}{l})\right]^{2}\] \[ = \frac{1}{2}\left[(M + m)\ddot x -
ml\dot\theta^{2}\sin\theta + mg\cos\theta\sin\theta - m\ddot
x\cos^{2}\theta\right]^{2}\; . \]

For  simplicity, denote by \[G^{\theta} = \frac{g\sin\theta - \ddot x\cos\theta}{l}.\]

Now, the presymplectic  $2$-form $\Omega_{W_0}$, the Hamiltonian ${H}_{W_0}$ and the primary  constraint $\varphi^1_{x}$ are, respectively
\begin{eqnarray*}
\Omega_{W_0} &=& dx\wedge dp_{x}^{0} + d\theta\wedge dp_{\theta}^{0} + d\dot x\wedge dp_{x}^{1} + d\dot\theta\wedge dp_{\theta}^{0}\; ,\\
{H}_{W_0} &=&p_{x}^{0}\dot x + p_{\theta}^{0}\dot\theta + p_{x}^{1}\ddot x + p_{\theta}^{1}\left[\frac{g\sin\theta - \ddot x\cos\theta}{l}\right] \\&&-
\frac{1}{2}\left[(M + m)\ddot x - ml\dot\theta^{2}\sin\theta +
mg\cos\theta\sin\theta - m\ddot x\cos^{2}\theta\right]^{2}\; ,
\\
\varphi^1_{x} &=& \frac{\partial\widetilde H}{\partial\ddot x} = p^{1}_{x} + p_{\theta}^{1}\frac{\partial G^{\theta}}{\partial\ddot x} - \frac{\partial\widetilde{L}_{{\mathcal M}}}{\partial\ddot x} = 0,\end{eqnarray*}
 i.e.,
\[p^{1}_{x}
 = - p_{\theta}^{1}\frac{\partial G^{\theta}}{\partial\ddot x} + \frac{\partial\widetilde{L}_{{\mathcal M}}}{\partial\ddot x}.\]

This constraint determines the submanifold $W_1$. Applying Proposition \ref{proposition} we deduce that the 2-form
 $\Omega_{W_1}$, restriction of $\Omega_{W_0}$ to $W_1$, is  symplectic since
\[
\frac{\partial^{2}\widetilde{L}_{{\mathcal M}}}{\partial\ddot x^2}  - p_{\theta}^{1}\frac{\partial^{2}G^{\theta}}{\partial\ddot x^2}= \left[(M + m)- m\cos^{2}\theta\right]^{2} \neq 0\; .
\]
%


Therefore, the
algorithm stabilizes at the first constraint submanifold $W_1$. Moreover, there exists a unique solution of the dynamics, the vector field  $X\in {\mathfrak X}(W_1)$ which satisfies $i_{X}\Omega_{W_{1}} = d {H}_{W_{1}}$.
 In consequence, we have a unique control input which extremizes (minimizes) the objective function ${\mathcal A}$ and then the force exerted to the car is the minimum possible.
If we take the flow $F_{t}: W_{1}\rightarrow W_{1}$ of the vector field $X$ then we have that $F_{t}^{*}\Omega_{W_1} = \Omega_{W_1}$.
Obviously, the hamiltonian function
\begin{eqnarray*}
\widetilde{H}\big{|}_{W_1} &=&
p_{x}^{0}\dot x + p_{\theta}^{0}\dot\theta + \left[ -
p_{\theta}^{1}\frac{\partial G^{\theta}}{\partial\ddot x} +
\frac{\partial\widetilde{L}_{N}}{\partial\ddot x}\right]\ddot x +
p_{\theta}^{1}\left[\frac{g\sin\theta - \ddot x\cos\theta}{l}\right]
-\\
&& \frac{1}{2}\left[(M + m)\ddot x - ml\sin\theta\dot\theta^{2} +
mg\cos\theta\sin\theta - m\ddot x\cos^{2}\theta\right]^{2}
\end{eqnarray*}
is preserved by the solution of the optimal control problem, that is  $\widetilde{H}\big{|}_{W_1}\circ F_t=\widetilde{H}\big{|}_{W_1}$.
Both  properties, symplecticity and preservation of energy, are important geometric invariants. In next section,  we will construct, using discrete variational calculus, numerical integrators which inherit some of the geometric
properties of the optimal control problem (symplecticity, momentum
preservation and, in consequence,  a very good energy behavior).
}

\end{example}
\section{Geometric discretization of optimal control problems for underactuated mechanical systems}\label{section6}

\subsection{Discrete vakonomic mechanics}
 In this section we discuss some ideas from discrete mechanics for  vakonomic systems (see \cite{benitodiego}).
 The main idea is to use discrete variational calculus. In the case of vakonomic systems, the  principle seeks to find a discrete curve which is
a critical point of the discrete action sum subject to some constraint functions.  The discrete curves are sequences of points that  approximate curves on $Q$.

 The discretizing procedure of a given continuous vakonomic system, determined by a Lagrangian function $L: TQ\to \R$ and a constraint submanifold ${\mathcal M}$ of $TQ$, starts first substituting the velocity phase space $TQ$ by the cartesian product of two copies of $Q$, $Q\times Q$. Secondly,  we discretize the continuous lagrangian and the constraint submanifold to a discrete lagrangian function $L_d: Q\times Q\to \R$ and a constraint submanifold ${\mathcal M}_d$ of $Q\times Q$ determined by the vanishing of $m$-independent constraints functions $\Phi_{d}^{\alpha}: Q\times Q\rightarrow \R$.

Given  $q_0$ and $q_N$ in $Q$,  for some (also fixed) integer $N$,  we consider the space
of sequences    $(q_0, q_1, \ldots, ,q_{N})$ (the  discrete paths joining $q_0$
to $q_N$). We form the discrete action sum
\[
\mathcal{A}_{d}(q_0,q_1,\ldots, q_N)=\sum_{k=0}^{N-1} L_{d}(q_{k}, q_{k+1})
\]
over discrete paths satisfying the discrete constraints equations, that is $\Phi_d^{\alpha}(q_k, q_{k+1})=0$, with $k=0, \ldots, N-1$.
We compute the critical point of this action sum subjected to the constraint equations; that is,
\begin{equation}\label{mnb-1}
 \left\{
   \begin{array}{l}
   \hbox{min }\mathcal{A}_{d}(q_0,q_1,\ldots, q_N)\hspace{1.75 cm}\hbox{with } q_0 \hbox{ and }q_N \hbox{ fixed}\\
   \hbox{suject to }\Phi^{\alpha}_d(q_k,q_{k+1})=0, \hspace{1.cm} 1\leq \alpha\leq m \hbox{ and }\ 0\leq k\leq
   N-1\; .
   \end{array}
 \right.
\end{equation}
Observe that system is subjected to $Nm$ constraint
functions.

We define the \textbf{augmented Lagrangian} ${\mathcal L}_d: Q\times Q\times \R^m\to \R$  by
$${\mathcal L}_d(x,y,\lambda) =
L_d(x,y)+\lambda_\alpha\Phi^\alpha_d(x,y).$$
This Lagrangian $\mathcal{L}_{d} $ gives rise to the following unconstrained discrete variational problem
\begin{equation}\label{mnb-2}
 \left\{
 \begin{array}{l}
  \hbox{min }
\overline{\mathcal{A}}_{d}\;(q_0, q_1,\ldots,q_N,\lambda^0,\lambda^1,\ldots,\lambda^{N-1})\;\hspace{.3 cm }\hbox{ with } q_0 \hbox{ and } q_N \hbox{ fixed }\; ,\\
 q_k\in Q,\hspace{.5 cm }\lambda_k\in \real^{m}\hspace{.5 cm }
k=0,\ldots,N-1,\hspace{.5 cm} q_N\in Q,
 \end{array}
\right.
\end{equation}
where
\begin{eqnarray*}\label{pvdsv}\;\overline{\mathcal{A}}_{d}\;(q_0,q_1,\ldots,q_N,\lambda^0,\lambda^1,\ldots,
\lambda^{N-1})&=&
\sum_{k=0}^{N-1}\;\mathcal{L}_d(q_k,q_{k+1},\lambda^{k})\\
 &=& \sum_{k=0}^{N-1}\;\left[{L}_d(q_k,q_{k+1})+
\lambda_{\alpha}^{k}\Phi^\alpha_d (q_k,q_{k+1})\right]\; ,
\end{eqnarray*}
 where
$\lambda^k$ is a $m$-vector with components $\lambda^k_{\alpha}$ with $
1\leq \alpha\leq m$.

From the classical lagrangian multiplier theorem, we have that the regular extremals of Problem \eqref{mnb-1} are the same than in Problem \eqref{mnb-2}.
  Therefore, applying standard discrete variational calculus we deduce that the solutions of problem \eqref{mnb-1} verify the following set of
  difference equations
%
\[
\left\{
\begin{array}{l}\label{lelo}
D_1L_d(q_k, q_{k+1})+D_2L_d(q_{k-1},q_k) + \lambda_{\alpha}^{k}
D_1\Phi^{\alpha}_d(q_k, q_{k+1}) + \lambda_{\alpha}^{k-1}
D_2\Phi^{\alpha}_d (q_{k-1}, q_k) = 0\; ,
{1\leq k\leq N-1} \, ,\\
\Phi^{\alpha}_d(q_k,q_{k+1}) = 0\; , \hspace{0.4 cm}1\leq \alpha\leq m
\; ,\hspace{0.1 cm} \hbox{ and }  0\leq k\leq N-1
\end{array}
\right.
\]
where $D_1 L_d$ and $D_2 L_d$ denote the derivatives of the
discrete lagrangian $L_d$ respect to the first and the second argument,
respectively.

For all function $F\in C^{\infty}(Q\times Q)$,  we denote by  $D_{12}F$ the $n\times n$-matrix
$\displaystyle{\left(\frac{\partial^2 F}{\partial x^i\partial
y^j}\right)}$ (partial derivatives with respect to the first and second variables).   Then, if the matrix  \[ \left(
\displaystyle{\begin{array}{cc}
\displaystyle{D_{12} L_d+\lambda_{\alpha}D_{12}\Phi^{\alpha}_d}& \displaystyle{\frac{\partial \Phi^{\alpha}_d}{\partial x}}\\
\displaystyle{\left(\frac{\partial\Phi^{\alpha}_d}{\partial
y}\right)^T}&{\mathbf 0}_{m\times m}
\end{array}}
\right)_{(n+m)\times (n+m)}
\]
is regular along ${\mathcal M}_d\times \R^m$, then, by a direct application of the implicit function theorem, we deduce that there exists a unique map  \[
\begin{array}{rrcl}
\Upsilon_d:& {\mathcal M}_d\times \R^m&\longrightarrow& {\mathcal M}_d\times \R^m\\
       & (x,y,\lambda)&\longmapsto&(y,v,\Lambda)\; ,
\end{array}
\]
such that for all solutions  $(q_0,
q_1,\ldots, q_N, \lambda^0, \lambda^1, \ldots, \lambda^{N-1})$ of
the equation $\eqref{lelo}$ we have that
$$
\Upsilon_d(q_{k-1},q_k,\lambda^{k-1})=(q_k,q_{k+1},\lambda^{k})\; .
$$
The application $\Upsilon_d$ will be called the \textbf{discrete flow} of the  vakonomic problem.

In \cite{benitoleondediego}, it is shown that the discrete flow $\Upsilon_d$ preserves a symplectic form naturally defined on ${\mathcal M}_d\times \R^m$ and it is momentum preserving if the discrete Lagrangian $L_d$ and the constraint submanifold ${\mathcal M}_d$ are invariant under the action of a Lie group of symmetries.

\subsection{Discrete second-order  vakonomic mechanics }\label{vako-dis}

In this subsection, we study  discrete second-order mechanical systems with constraints (see \cite{benitoleondediego}) bearing in mind  the discretization of optimal control problems for underactuated mechanical systems.

A natural discrete space substituting the second-order tangent bundle $T^{(2)}Q$ is $Q\times Q\times Q$
and therefore a discrete vakonomic system is determined by a discrete lagrangian $\tilde{L}_d: Q\times Q\times Q\longrightarrow \R$
 and a constraint submanifold ${\mathcal M}_d$ is locally determined by the vanishing of $m$-constraint functions $\Phi_d^{\alpha}: Q\times Q\times Q\ra \R$.
%


 Given a discrete lagrangian $\tilde{L}_d: Q\times Q\times Q\longrightarrow \R$ we define the \emph{discrete action} $\mathcal{A}_{d}: Q^{N+1}\ra\R$ by \[\mathcal{A}_{d}(q_{0},...,q_{N})=\sum_{k=0}^{N-2}\tilde{L}_{d}(q_{k}, q_{k+1}, q_{k+2}).\]

 Adding the constraints, we have the following discrete constrained variational problem  \begin{equation}\label{seco-vako}
 \left\{
 \begin{array}{l}
 \hbox{ min }\mathcal{A}_{d}(q_{0}, q_{1},..., q_{N}), \hspace{1.75 cm}\hbox{with } q_0, q_1 \hbox{ and }q_{N-1}, q_N \hbox{ fixed}\\\\
 \hbox{subject to } \Phi^{\alpha}_{d}(q_{k}, q_{k+1}, q_{k+2})=0\; , \hbox{ with } 1\leq\alpha\leq m,\ 0\leq\ k\leq N-1.
\end{array}
\right.
\end{equation}

As in the previous section we define the augmented  lagrangian ${\mathcal  L}_{d}: Q\times Q\times Q\times\R^{m}\ra\R$
 given by $${\mathcal  L}_{d}(q_{k}, q_{k+1}, q_{k+2}, \lambda^k) = \widetilde{L}_{d}(q_{k}, q_{k+1}, q_{k+2}) +
   \lambda^k_{\alpha}\Phi^{\alpha}_{d}(q_{k}, q_{k+1}, q_{k+2})$$ with
  $q_{k}\in Q; \ \lambda^{k}=(\lambda^k_1, \ldots, \lambda^k_m)\in \R^{m}, \ 0\leq k\leq N-2$.
Hence, as in the previous section, Problem (\ref{seco-vako}) is equivalent to the following (singular) unconstrained problem for ${\mathcal L}_d$
\begin{equation}\label{seco-vako-2}
 \left\{
 \begin{array}{l}
  \hbox{min }
\overline{\mathcal{A}}_{d}\;(q_0, q_1,\ldots,q_N,\lambda^0,\lambda^1,\ldots,\lambda^{N-1})\;\hspace{.3 cm }\hbox{ with } q_0, q_1 \hbox{ and } q_{N-1}, q_N \hbox{ fixed }\; ,\\
 (q_k, q_{k+1}, q_{k+2})\in Q\times Q\times Q,\hspace{.5 cm }\lambda_k\in \real^{m}\; ,\hspace{.5 cm }
k=0,\ldots,N-2,
 \end{array}
\right.
\end{equation}
where
\begin{eqnarray*}\label{pvdsv}\;\overline{\mathcal{A}}_{d}\;(q_0,q_1,\ldots,q_N,\lambda^0,\lambda^1,\ldots,
\lambda^{N-2})&=&
\sum_{k=0}^{N-2}\;\mathcal{L}_d(q_k,q_{k+1},q_{k+2},\lambda^{k})\\
 &=& \sum_{k=0}^{N-2}\;\left[\widetilde{L}_d(q_k,q_{k+1}, q_{k+2})+
\lambda_{\alpha}^{k}\Phi^\alpha_d (q_k,q_{k+1}, q_{k+2})\right]
\end{eqnarray*}
 and
$\lambda^k$ is a $m$-vector with components $\lambda^k_{\alpha}$ , $
1\leq \alpha\leq m$.

 Hence, the extremality conditions are
\begin{eqnarray*}
0& =& D_{3}\tilde{L}_{d}(q_{k-2}, q_{k-1}, q_{k}) + D_{2}\tilde{L}_{d}(q_{k-1}, q_{k}, q_{k+1}) \\
&&+ D_{1}\tilde{L}_{d}(q_{k}, q_{k+1}, q_{k+2})+ \lambda^{k-2}_{\alpha}D_{3}\Phi_d^{\alpha}(q_{k-2}, q_{k-1}, q_{k})\\
&&
+ \lambda^{k-1}_{\alpha}D_{2}\Phi_d^{\alpha}(q_{k-1}, q_{k}, q_{k+1}) +
\lambda^k_{\alpha}D_{1}\Phi^{\alpha}_d(q_{k}, q_{k+1}, q_{k+2}), \\
0&=&\Phi_d^{\alpha}(q_{k-2}, q_{k-1}, q_{k}) \\
0&=&\Phi_d^{\alpha}(q_{k-1}, q_{k}, q_{k+1}) \\
0&=&\Phi_d^{\alpha}(q_{k}, q_{k+1}, q_{k+2})\; .
\end{eqnarray*}
where $2\leq k \leq N-2$.

If the matrix \begin{equation}\label{regu}
\det\left(
\begin{array}{cc}
 D_{1 3}\tilde{L}_{d}(x, y, z) + \lambda_{\alpha}D_{1 3}\Phi_d^{\alpha}(x, y, z) & D_{3}\Phi_d^{\alpha}(x, y, z) \\
D_{1}\Phi_d^{\alpha}(x, y, z) & 0 \\
\end{array}
\right)\neq 0,
\end{equation} is regular for all $(x, y, z)\in {\mathcal M}_d=\{(x,y,z)\in Q\times Q\times Q\; \big|\; \Phi_d^{\alpha}(x,y,z)=0\}$ and $\lambda_{\alpha}\in \R$, $1\leq \alpha\leq m$.
 Assuming this regularity assumption and  by a direct application of the implicit function theorem, we deduce that there exists a unique application
 \[
\begin{array}{rrcl}
\Upsilon_d:&\overline{\mathcal M}_d\times \R^{2m}&\longrightarrow&\overline {\mathcal M}_d\times \R^{2m}\\
    & (q_0, q_1, q_2, q_3,\lambda^0_{\alpha}, \lambda^1_{\alpha})&\longmapsto& (q_1, q_2, q_3, q_4,\lambda^1_{\alpha}, \lambda^2_{\alpha})
\end{array}
\]
which univocally determines $q_4$ and $\lambda^2_{\alpha}$, $1\leq \alpha\leq m$ from the initial conditions  $(q_0, q_1, q_2, q_3,\lambda^0_{\alpha}, \lambda^1_{\alpha})$.
Here, $\overline{\mathcal M}_d$ denotes the submanifold of $Q^4=Q\times Q\times Q\times Q$
\[
\overline{\mathcal M}_d=\{ (q_0, q_1, q_2, q_3)\in Q^4\; \big|\; \Phi^{\alpha}_d(q_0,q_1, q_2)=0, \Phi^{\alpha}_d(q_1,q_2, q_3)=0, 1\leq \alpha\leq m\}\; .
\]
The mapping $\Upsilon_d$
will be called the  \textbf{discrete second-order vakonomic flow}.

Using similar techniques than in \cite{benitodiego,benitoleondediego}, it is possible to show that, under the regularity assumptions,  the discrete second-order vakonomic flow is symplectic and preserves momentum in the case when we have a Lie group action  preserving the discrete lagrangian $L_d$ and the constraint submanifold ${\mathcal M}_d$.

\subsection{Application to optimal control of underactuated
mechanical systems}

In this section, we  show that the discrete vakonomic approach of second order problem is an appropriate framework for discrete versions of optimal control problems of  underactuated mechanical systems considered in Section \ref{section5} (see \cite{objuma} for an alternative approach). The main application will be the explicit  construction of geometric numerical integrators for this type of systems.

Let us take a discrete lagrangian $L_d: Q\times Q\to \R$ where $Q=Q_1\times Q_2$ as in Section \ref{section5}.
Then, an element $(q_0^A, q_1^A)\in Q\times Q$ admits a global decomposition of the form $(q_0^a, q_0^\alpha, q_1^a, q_1^\alpha)$. Thus, we can consider the following
\textbf{discrete underactuated mechanical  system}
\begin{eqnarray*}\label{aqr}
D_{2}^{a}L_{d}(q_{k-1}^{A}, q_{k}^{A}) + D_{1}^{a}L_{d}(q_{k}^{A},
q_{k+1}^{A}) &=& u_k^{a}\\
D_{2}^{\alpha}L_{d}(q_{k-1}^{A}, q_{k}^{A}) + D_{1}^{\alpha}L_{d}(q_{k}^{A}, q_{k+1}^{A}) &=& 0\; ,
\end{eqnarray*}
$1\leq A\leq n$, $1\leq a\leq m$, $m+1\leq \alpha\leq n$.
Here $D_{i}^{a}L_{d}$ and $D_{i}^{\alpha}L_{d}$ represent the partial
derivatives with respect to  coordinates $a$ and $\alpha$,
respectively.

The optimal control problem is determined prescribing the discrete cost functional
$$\mathcal{A}_{d} = \sum_{k=1}^{N-1} C(q^A_{k}, q^A_{k+1}, u_k^{a})$$ with initial and final conditions
 $q_0, q_1$ and  $q_{N-1}, q_N$, respectively.

Since the control variables appear explicitly the previous discrete
optimal control problem is equivalent to the  second-order discrete vakonomic problem determined by
\begin{eqnarray*}
\widetilde{L}_d(q_{k-1}^{A},  q_{k}^{A}, q_{k+1}^{A})
&=& C\left(q_{k}^{A},  q_{k+1}^{A},  D_{2}^{a}L_{d}(q_{k-1}^{A}, q_{k}^{A}) + D_{1}^{a}L_{d}(q_{k}^{A},
q_{k+1}^{A})\right)\\
\Phi_d^\alpha (q_{k-1}^{A},  q_{k+1}^{A},  q_{k+2}^{A})&=&D_{2}^{\alpha}L_{d}(q_{k-1}^{A}, q_{k}^{A}) + D_{1}^{\alpha}L_{d}(q_{k}^{A}, q_{k+1}^{A})=0\; .
\end{eqnarray*}

Applying the techniques developed in Subsection \ref{vako-dis} and assuming  the regularity condition  (\ref{regu})
we obtain the discrete flow
\[
\Upsilon_d:\overline{\mathcal M}_d\times \R^{2m}\longrightarrow\overline {\mathcal M}_d\times \R^{2m}\; .
\]

\begin{example} \textbf{The Discrete Cart-Pole System: }\
{\rm
(See Example \ref{cart-pole})

Consider the following  discrete Lagrangian  $L_{d}:Q\times Q\ra\R$ where $Q =
\R\times\mathbb{S}^{1}$
\begin{eqnarray*}
&&L_{d}(x_{k-1}, \theta_{k-1}, \ x_k, \theta_k) = \frac{1}{2}M\left(\frac{x_k - x_{k-1}}{h}\right)^{2}\\
&&+\frac{1}{2}m\left[\left(\frac{x_k - x_{k-1}}{h}\right)^{2} + 2\left(\frac{x_k - x_{k-1}}{h}\right)l\cos\left(\frac{\theta_k + \theta_{k-1}}{2}\right)\left(\frac{\theta_k - \theta_{k-1}}{h}\right)
+ l^{2}\left(\frac{\theta_k - \theta_{k-1}}{h}\right)^{2}\right]\\
&&- mgl\cos\left(\frac{\theta_k + \theta_{k-1}}{2}\right) - mg\widetilde{h}
\end{eqnarray*}
where $\widetilde{h}$ is the car height.

The discreted controlled Euler-Lagrange equations are
\begin{eqnarray*}
 &&(2x_{k+1} - x_k - x_{k+2})(\frac{M+m}{h^{2}})\\
  &&+\frac{ml}{h^2}\left[\cos\left(\frac{\theta_{k+1} +
\theta_k}{2}\right)\left(\theta_{k+1} - \theta_k\right) -
\cos\left(\frac{\theta_{k+2} +
\theta_{k+1}}{2}\right)\left(\theta_{k+2} -
\theta_{k+1}\right)\right] = u_{k}\\
&&\frac{l^{2}m}{h^{2}}(2\theta_{k+1}-\theta_{k}-\theta_{k+2}) + \frac{lm}{h^{2}}\left[(x_{k+1}-x_{k})\cos\left(\frac{\theta_{k+1}+
\theta_{k}}{2}\right) -
(x_{k+2}-x_{k+1})\cos\left(\frac{\theta_{k+2}+\theta_{k+1}}{2}\right)\right]
\\
&&+\frac{lmg}{2}\left[\sin\left(\frac{\theta_{k+1}+\theta_{k}}{2}\right)
+ \sin\left(\frac{\theta_{k+2} + \theta_{k+1}}{2}\right)\right] \\
&&-\frac{lm}{2h^{2}}\left[(x_{k+1} - x_k)(\theta_{k+1} -
\theta_k)\sin\left(\frac{\theta_{k+1} + \theta_k}{2}\right) +
(x_{k+2} - x_{k+1})(\theta_{k+2} -
\theta_{k+1})\sin\left(\frac{\theta_{k+2} +
\theta_{k+1}}{2}\right)\right] = 0.
\end{eqnarray*}

For solving the associated  discrete optimal control problem, we need to find sequences  $\{(q_k,
q_{k+1}, u_k)\}$, minimizing the cost functional $$\mathcal{A}_{d}
=\frac{1}{2} \sum_{k=1}^{N-1} u_{k}^{2}.$$ We know that this problem
is equivalent to solve the following variational problem with
constraints \[\min \widetilde{\mathcal{A}}_{d}
=\sum_{k=0}^{N-2}\widetilde{L}(x_{k},\theta_{k},x_{k+1},\theta_{k+1},x_{k+2},\theta_{k+2})\]subject
to the constraints
\begin{eqnarray*}
&&\Phi_{d}(q_k, q_{k+1}, q_{k+2}) = \\
&&l^{2}m(2\theta_{k+1}-\theta_{k}-\theta_{k+2}) + lm\left[(x_{k+1}-x_{k})\cos\left(\frac{\theta_{k+1}+
\theta_{k}}{2}\right) -
(x_{k+2}-x_{k+1})\cos\left(\frac{\theta_{k+2}+\theta_{k+1}}{2}\right)\right]
\\
&&+\frac{lmgh^2}{2}\left[\sin\left(\frac{\theta_{k+1}+\theta_{k}}{2}\right)
+ \sin\left(\frac{\theta_{k+2} + \theta_{k+1}}{2}\right)\right]\\&& -
\frac{lm}{2}\left[(x_{k+1} - x_k)(\theta_{k+1} -
\theta_k)\sin\left(\frac{\theta_{k+1} + \theta_k}{2}\right) +
(x_{k+2} - x_{k+1})(\theta_{k+2} -
\theta_{k+1})\sin\left(\frac{\theta_{k+2} +
\theta_{k+1}}{2}\right)\right] = 0,
\end{eqnarray*}
 with $k=0,\ldots, N-2$ where
$\widetilde{L}_{d}:Q\times Q\times Q\ra\real$ is given by
\begin{eqnarray*}
\widetilde{L}_{d}(q_{k}, q_{k+1}, q_{k+2}) &=& \frac{1}{2}u_{k}^{2}\\
&=&\frac{(M+m)^2}{2h^{4}}(2x_{k+1}-x_k-x_{k+2})^2 \\
&&+ \frac{{(ml)}^{2}}{2h^4}\left[\cos\left(\frac{\theta_{k+1}+\theta_{k}}{2}\right)
\left(\theta_{k+1}-\theta_{k}\right)-\cos\left(\frac{\theta_{k+2}+\theta_{k+1}}{2}\right)\left(\theta_{k+2}-\theta_{k+1}\right)\right]^{2}
\end{eqnarray*}
To test our numerical algorithm we have programmed it in Matlab. The
program admits as input data $(q_0, q_1, q_2, q_3,\lambda_{0}, \lambda_{1})$ and the number of steps  $N$; then it computes $q_4$ and $\lambda_{2}$ and
replace the initial data with $(q_1, q_2,  q_3,  q_4, \lambda_{1}, \lambda_{2})$ to calculate $q_5$ and $\lambda_{3}$,  etc.
\newpage
In the figure  we show the excellent energy behavior (of the function $H_{W_1}$) of our symplectic geometric integrator for the cart-pole system
\begin{center}
\includegraphics[width=15cm]{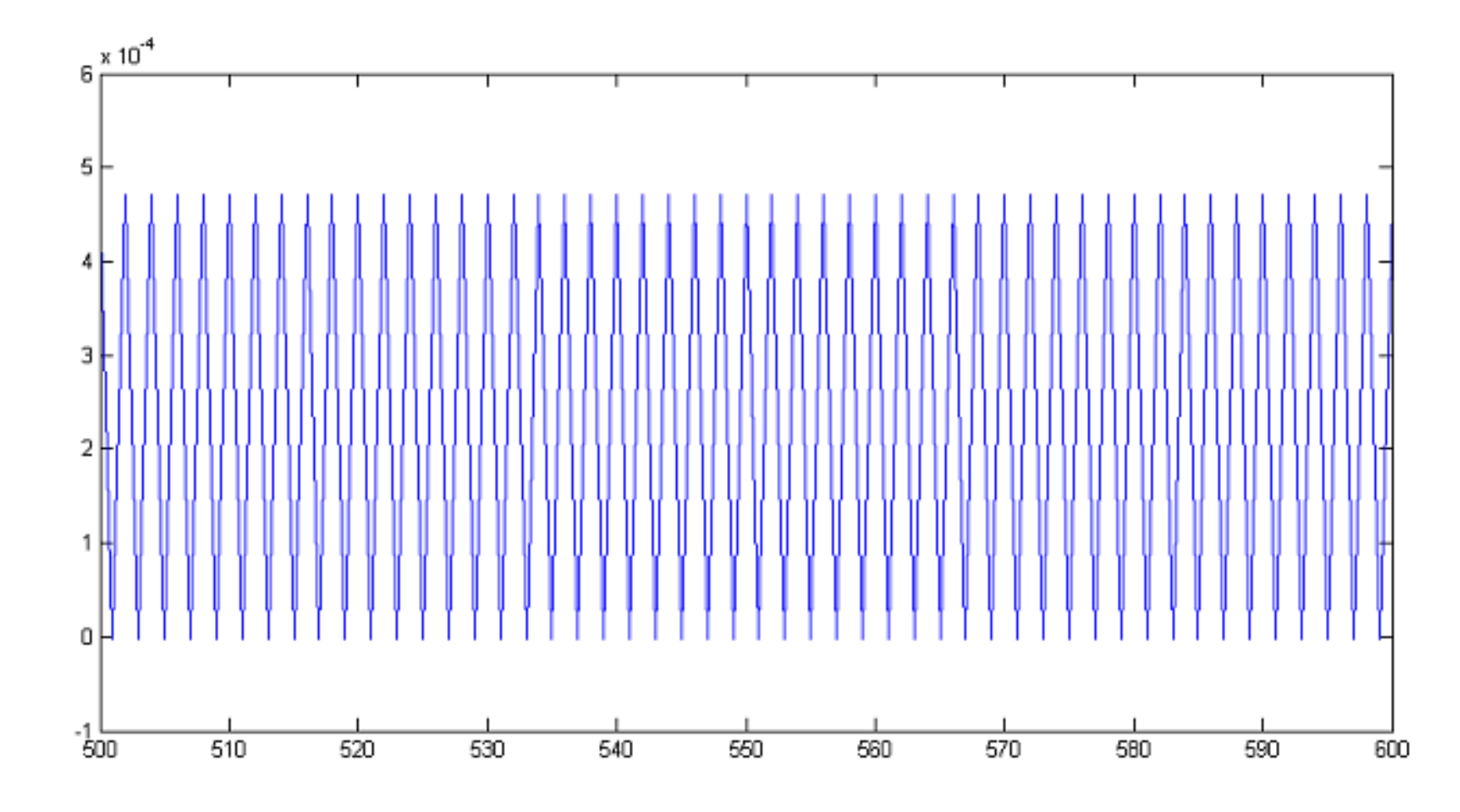}
\end{center}

}
\end{example}


\end{document}